# Bearings Degradation Monitoring Indicators Based on Discarded Projected Space Information and Piecewise Linear Representation

**Fei Huang***

Laboratoire de Conception, Optimisation et Modélisation des Systèmes, LCOMS EA 7306
Université de Lorraine, Metz 57000, France
and
Huaiyin Institute of Technology
1 Meicheng Rd, Huaian 223003, Jiangsu, P.R. China
Email: fei.huang@univ-lorraine.fr
*Corresponding author

**Alexandre Sava**

Laboratoire de Conception, Optimisation et Modélisation des Systèmes, LCOMS EA 7306
Université de Lorraine, Metz 57000, France
Email: alexandre.sava@univ-lorraine.fr

**Kondo H. Adjallah**

Laboratoire de Conception, Optimisation et Modélisation des Systèmes, LCOMS EA 7306
Université de Lorraine, Metz 57000, France
Email: kondo.adjallah@univ-lorraine.fr

**Zhouhang Wang**

Laboratoire de Conception, Optimisation et Modélisation des Systèmes, LCOMS EA 7306
Université de Lorraine, Metz 57000, France
Email: zhouhang.wang@univ-lorraine.fr

**Biographical notes:** Fei Huang is currently pursuing his PhD at Université de Lorraine, France. He received his MS from Jiangsu University, China in 2009 and BS from Tianjin University of Science & Technology, China in 2003. His research is focused on health condition monitoring and decision support for predictive maintenance.

Alexandre Sava received his PhD from Institut National Polytechnique de Grenoble, France in 2002. Currently, he is an associate professor at Université de Lorraine and senior researcher at LCOMS laboratory, France. His main research topic concerns the monitoring and health management of durable infrastructures.

Kondo H. Adjallah received his PhD from Institut Nationale Polytechnique de Lorraine, France.  Currently, he is a distinguished Professor (Chevalier de l'Ordre des Palmes Académiques) in France. He contributed in cutting edge research in leading organizations including the CRAN of Nancy, Charles Delaunay Institute of Troyes. He was an Associate Professor (1994-2008) at UTT (Université de Technologie de Troyes), France, then a full Professor at the Université de Lorraine with the LCOMS research unit. His current research interests include data integration and analysis, reliability and degradation modeling, health condition monitoring and decision support for predictive maintenance. He is the member of the IEEE society since 2002.

Zhouhang Wang received his PhD from Université de Lorraine, France in 2013. He obtained his MS in Mechanical Engineerings in 2008 and Engineer diploma in 2007 from the Ecole Nationale d'Ingénieurs de Metz, France. His research interests include reliability and degradation modeling, health condition monitoring and decision support for predictive maintenance.

**Abstract:** Condition-based maintenance of rotating mechanics requests efficient bearings degradation monitoring. The accuracy of bearings degradation measure depends largely on degradation indicators. To extract efficient indicators, in this paper we propose a method based on the discarded projected space information and piecewise linear representation (PLR) to build three bearings degradation monitoring indicators which are named $SDHT^2$, $VSDHT^2$ and $NVSDHT^2$. The discarded projected space information is measured by the segmented discarded Hotelling T square we propose in this paper. For illustration, the IEEE PHM 2012 benchmark data set is used in this paper. The results show that the three new indicators are all sensitive and monotonic during the bearings whole lifecycle. They describe the whole degradation process history and carry the real-time information of bearings degradation. And $NVSDHT^2$ is the generalized version of $VSDHT^2$, which is promising to monitor bearings degradation.

## 1 Introduction

To prevent the breakdowns in manufacturing systems, condition-based maintenance (CBM) is an efficient policy for modern industries (Jardine et al., 2006). Bearings are common components in rotary machinery, bearing failures can severely influence the overall systems performance (Nandi et al., 2005). Costly breakdowns during manufacturing processes can be resulted in by the unexpected bearing failure. Hence, bearings degradation monitoring is crucial for condition-based maintenance of rotating mechanics.

In general, the bearings health condition can be revealed using vibration signal data. Thus, to monitor bearings degradation accurately, one can start by extracting key degradation features from the bearings vibration signal. Vibration signals based common features may include the mean value, the standard deviation (STD), the root mean square (RMS), the root-square amplitude, the skewness, the peak to peak, the waveform index, the pulse index, the margin index, the kurtosis and so on (Yu et al., 2016).

For bearings such mechanical components, there is no self-repair during periods of non-utilization. Therefore, monotonic features can well characterize bearings degradation evolution. Unfortunately, the aforementioned common features are not monotonic during the degradation process. They are only sensitive for specific stages of the degradation process. During bearings lifetime, each of these features considered separately is not efficient for the degradation monitoring.

To cope with this situation, authors devoted to building new indicators that can characterize better the bearings degradation processes, compared to the common features based on the vibration signal (Benkedjouh et al., 2013; Rai et al., 2017; Ahmad et al., 2017; Liu et al., 2017; Mahamad et al., 2010; Ali et al., 2015). On one hand, the indicators proposed in existing articles are devoted to monitor the last part of the lifetime when the degradation becomes significant. On the other hand, these indicators are associated with the current degradation state, and provide no information on the historical evolution of the degradation process. Bearings degradation processes are complex and do not only depend on the current degradation features but also on what happened in the past on the bearings. Therefore, we believe that it is valuable for a bearing degradation indicator to be sensitive and monotonic during the whole degradation process and reveal not only the current degradation but also information on the historical degradation process. To the best of our knowledge, no existing bearing degradation indicator relies on memorizing historical evolution of the degradation process.

In Zhang et al. (2006), authors have extracted composite indicators from different common features based on a vibration signal using the principal component analysis method (PCA). In Dong et al. (2013), PCA is used to merge the common features extracted from the vibration signal to reduce the features space dimension to obtain sensitive features. In Zhang et al. (2006) and Dong et al. (2013), vibration signal based common features are projected into a specific space. Then, new indicators are obtained in a lower dimension subspace. However, discarded components may also carry useful information (Rencher et al., 2003). Moreover, in some applications, the information in the rejected indicators space might be more useful than the information carried by those in the reduced space (Rencher et al., 2003). Hence, we propose in this paper to build indicators based on the information in the discarded space. This corresponds to the information lost by the projection on the principal components. We use the segmented discarded Hotelling T square described in section 3 for quantifying the difference between the reduced projected space and the full projected space to monitoring bearings degradation. The indicator $SDHT^2$ is built based on segmented discarded Hotelling T square. For tracking and memorizing the bearings degradation process evolution, we propose to combine segmented discarded Hotelling T square with a PLR approach to get indicators $VSDHT^2$ and $NVSDHT^2$. The effectiveness of the three new indicators for characterizing the degradation during entire bearings lifetime is tested through numerical experiments, using IEEE PHM 2012 data set from the PRONOSTIA test bed (Nectoux et al., 2012).

In the next Section, we will investigate the efficiency of existing features extracted from vibration signals using data from benchmark data. The following, Section 3 presents the indicator SDHT$^2$, while the indicator VSDHT$^2$ is presented in Section 4, and Section 5 proposes the last indicator NVSDHT$^2$. The numerical results obtained with the indicators presented based on benchmark data is in the section 6. Finally, Section 7 concludes this work and proposes some further directions.

## 2  Existing common features

In order to monitor the degradation process of bearings, in this paper we use six common time domain features extracted from the vibration signal. The RMS, derived from the vibration signals, provides additional information on the energy quantity of the signal. It is one important feature used to estimate the bearing degradation (Rao et al., 2011). Furthermore, we used the five following features sensitive to the degradations: mean, STD, peak-magnitude-to-RMS level (PMR), kurtosis and skewness. The STD quantifies the signal's average variation. The PMR quantifies the pluses intensity of the signal. The kurtosis reveals the degree of deviation of the signal distribution from the average and the skewness indicates the signal distribution asymmetry.

2.1 The benchmark data

The effectiveness of the indicator introduced in this paper for bearing degradation monitoring is tested using the IEEE PHM 2012 data set from the PRONOSTIA test bed (Nectoux et al., 2012).

This data set gathers data related to identical bearings subject to different load conditions. For our study, we have chosen two training bearings with the tags (1-1, 1-2) submitted to the load condition 1, and five test bearings (1-3, 1-4, 1-5, 1-6, 1-7) submitted to the same load condition.

The bearings under condition 1 operated at 1800 rpm with 4000 N radial load. Vibration signal data were collected on each of the bearings over a complete run-to-failure period. Two accelerometers allowed measuring the vibrations in the vertical and horizontal directions.

2.2 Features extracted from benchmark data

Data were sampled at 25.6 kHz sampling rate and 0.1s recording duration with 10s sample intervals. Thus, each sample interval is made of 2560 measures. The parameters of sample are summarized in Figure 1.

For simplifying, in this paper we consider only the vibration signals on horizontal direction. For example, in Figure 2 we draw the RMS evolution graph over the whole lifecycle of the bearing 1-2. One can see that this evolution is not monotonic over the bearing lifecycle of the bearing except on the last period where it increases quickly toward failure. The other five selected common features, have the same drawback regarding the sensitivity to different stages of the degradation.

The value of each of the six selected common features previously mentioned is calculated for each sample interval consisting of 2560 measures. These values are the entries of a features vector $fc_{int}$, where $int$ identifies the sample intervals (1). The evolution of the features vector defines a time series that we use to generate the VSDHT$^2$ indicator. The value of $fc_{int}$ associated to a sample interval is referred as observation.

$$fc_{\text{int}} = \begin{bmatrix} x_{RMS_{\text{int}}} & x_{mean_{\text{int}}} & x_{STD_{\text{int}}} & x_{PMR_{\text{int}}} & x_{kurtosis_{\text{int}}} & x_{skewness_{\text{int}}} \end{bmatrix}^T \qquad (1)$$

## 3 The SDHT$^2$ indicator

Considering the limitations of the six common features mentioned in the previous section for monitoring the bearings degradation, a valuable characteristic we request for the new indicator is monotonicity through the entire bearing lifecycle. This requirement is natural, knowing that bearings are mechanical components and self-repair is improbable during the non-utilization periods.

3.1 Discarded projected space information

The bearings degradation influences the underlying correlation structure of the selected common features $fc_{int}$. Therefore, we suggest using the underlying correlation structures of the selected common features $fc_{int}$ to track the bearings degradation. Multivariate analyses methods can serve to characterize these diverse underlying correlation structures (Rencher et al., 2003).

PCA method is the most commonly used multivariate analysis method, which identifies underlying principal components among a set of variables and help understanding the relations among variables. Through the PCA method, the data in the original space can be projected by orthogonal transformation into a new space with the same variable dimension. The new space is the full projected space. The data in the projected space is the projected space data called principal components.

The retained principal components embody most of the variance of the original data (Rencher et al., 2003). They are associated with the reduced projected space. In this work, we employed the cumulative percent variance to fix the number of principal components retained to effectively describe the data (Valle et al., 1999).

Commonly, retained principal components with the higher variances are used to describe the original data and reduce the space dimension for the data analysis. However, the discarded components with smaller variances may also carry useful information in some extent (Rencher et al., 2003). Hence, discarded Hotelling T square is used to estimate the difference between the full projected space and the reduced projected space to track the bearings degradation.

Figure 3 depicts the relationships among the original space, the full projected space and the reduced projected space through the PCA method, where $fc$ is the original space data, $\varphi$ is the projected space data, $g=1, 2, ..., E$ is the variable label of the original space data, $E$ is the dimension of the original space, $v=1, 2, ... B$ is the observations label in a segment, $B$ is the number of observations in a segment, $d=1, 2, ..., Z$ is the segments label, $Z$ denotes the number of segments. And $p$ is the number of retained principal components.

3.2 Segmented discarded Hotelling T square

The Hotelling T square is the measure of the multivariate distance between each observation and the mean of the data set. The squared Mahalanobis distance is employed to estimate the Hotelling T square (Johnson et al., 2002; Krzanowski et al., 2000; Seber et al., 1984; Jackson et al., 2005).

Comparing with Euclidean distance, Mahalanobis distance measurement also considers the variance and covariance of the variables when calculating the distances, which has the advantage that the principal components are equally weighted during the calculation. One can capture all changes that happen in the components even with small variances (De Maesschalck et al., 2000).

The Hotelling T square are defined in (2) as it follows

$$T_{d,v}^2 = (\varphi_{d,v}^{1\sim\beta} - \vartheta_d)\phi_d^{\beta^{-1}}(\varphi_{d,v}^{1\sim\beta} - \vartheta_d)^T \qquad (2)$$

with

$$\varphi_{d,v}^{1\sim\beta} = \left[\varphi_{d,v}^1, \varphi_{d,v}^2, \cdots, \varphi_{d,v}^\beta\right] \qquad (3)$$

$$\vartheta_d = \left[\vartheta_d^1, \vartheta_d^2, \cdots, \vartheta_d^\beta\right] \qquad (4)$$

(The definitions of superscripts and subscripts have been mentioned in section 3.1.) where $\phi_d^\beta$ is the covariance matrix of the first $\beta$ principal components in the $d^{th}$ segment, $\varphi_{d,v}^g$ is the $v^{th}$ observation value of the $g^{th}$ principal component in the $d^{th}$ segment, $\vartheta_d^g$ is the mean of the $g^{th}$ principal component in the $d^{th}$ segment.

When $\beta$ is equal to the variable dimension E of the original data, $T^2_{d,v}$ in formula (2) is the full projected space Hotelling T square $TF_{d,v}$. When $\beta$ is equal to $p$, $T^2_{d,v}$ in formula (2) is the reduced projected space Hotelling T square $TR_{d,v}$.

The discarded Hotelling T square $TD_{d,v}$ is defined in equation (5). The segmented discarded Hotelling T square $TD_d^{segmented}$, introduced in this paper, is the Discarded Hotelling T square value of a time series segment, defined in equation (6).

$$TD_{d,v} = TF_{d,v} - TR_{d,v} \qquad (5)$$

$$TD_d^{segmented} = \frac{1}{B}\sum_{v=1}^{B} TD_{d,v} \qquad (6)$$

3.3 The indicator SDHT$^2$

As bearings degradation is a complex process. We believe that it is valuable for a bearing degradation indicator to carry information on the historical degradation process. Therefore, we think that enhancing the indicator by considering the information on the evolution of the bearings degradation process will increase its effectiveness.

Since $TD_d^{segmented}$ can characterize a segment of the time series $fc$. Characterizing all the historical data of the time series $fc$, the value of $TD_d^{segmented}$ is the value of the indicator SDHT$^2$ corresponding to the current time. In other words, we use $TD_d^{segmented}$ to survey the multivariate distance in the projection space to track the bearings degradation. The numerical results are showed in Section 6.

**4 The VSDHT² indicator**

The indicator SDHT² presented in previous section carries the information of the evolution of the bearings degradation process. It treats the historical process as a single block without considering the historical process has multiple stages.

If one considers that the historical process has multiple stages, we need to distinguish each stage of the historical process then use $TD_d^{segmented}$ to respectively characterize each stage of the historical process. Hence, we propose using a piecewise linear approximation method PLR of time series to divide the historical data of the time series *fc* into homogeneous segments (Keogh et al., 2001) corresponding to the stages of the historical process.

To implement the PLR approach, we divided the time series data *fc* into homogeneous segments. Then we calculated the feature over each segment. Each homogeneous segment corresponds to a certain stage of the bearing degradation process. The VSDHT² indicator calculated at a given time point, incorporates the degradation process information of the bearing on all previous stages.

We used a specific segmentation method to define time segments of the time series *fc*. For the feature evaluation, $TD_d^{segmented}$ values are respectively used as the characteristic values associated to each segment. Numerical results are given in Section 6.

4.1 Segmentation algorithms

Several algorithms are proposed in the literature for segmenting time series. Normally, one can classify them into three groups: Sliding Windows, Top-Down and Bottom-Up algorithms (Keogh et al., 2001).

Each group of segmentation algorithms has its own specific advantages and disadvantages. The Sliding Windows algorithms are not able to divide a time series data into predefined number of segments. Top-Down algorithms are more complex than the Bottom-Up algorithms (Keogh et al., 2001).

To build indicators from time series data of multi-features, we have segmented the time series into homogeneous data segments related to different degradation stages. As a requirement, the indicators must have the same dimension.

A Bottom-Up algorithm seems to be more suitable for our investigations because of a lower complexity and the possibility to fix the desired final number of segments.

Since Hotelling T square can be used to segment the time series data into homogeneous sections (Hotelling et al., 1933), we suggest using the $TD_d^{segmented}$ function (6) as the cost to segment the time series data into homogeneous parts.

4.2 The VSDHT² indicator

First, we use the segmentation method to divide the time series data into homogeneous segments which are associated with the stages of the bearings degradation process. The time series data corresponds to the time evolution of the features vector *fc* as mentioned in Section

2. Then, the characteristic values $TD_d^{segmented}$ from each segment are calculated respectively. $TD_d^{segmented}$ is also used as the cost value in the segmentation algorithm to divide the time series data into homogeneous segments.

4.2.1 The Extraction Algorithms

The VSDHT$^2$ health indicator is defined in equation (7). The pseudo-code of the Bottom-Up algorithm for extracting the VSDHT$^2$ is presented hereafter.

Note that: when calculating the $TD_d^{segmented}$ as the cost value, the labels $d$ and $v$ in formula (2-6) are associated with the segments consisting of adjacent segments. When calculating the $TD_d^{segmented}$ as the characteristic value, the labels $d$ and $v$ in formulas (2-6) are associated with final segments of the time series.

$$VSDHT^2 = \left[ TD_1^{segmented}, TD_2^{segmented}, \cdots, TD_m^{segmented} \right]^T \qquad (7)$$

---

**Algorithm 1: VSDHT$^2$ extraction**

1. Let $l$ be the length of the selected common features time series $fc$.
2. Fix the final number of segments $m$ and the parameter $k$ used to define the initial segments of the time series $fc$.
3. Divide the observations into segments such that the length of each segment except the last one is $ceil(l/k)$ (the round up value of $(l/k)$) and the length of the last segment is the remainder of $(l/ceil(l/k))$.
   where $l > k > m$
4. While $k>m$,
   Combine every two adjacent segments and calculate the cost value using (6);
   Merge the two adjacent segments with the smallest cost value;
   $k=k-1$;
   end (while).
5. Calculate the characteristic value of each segment using (6).
The vector with $m$ entries is VSDHT$^2$.

---

The setting of the parameters $m$ and $k$ of the algorithm 1 are discussed in the last part of this section.

Let us consider a time series $fc$, $J$ is the length of a segment and it is a constant. The integer $int$ varies from $0$ to $J$. This time series is divided into $m$ homogeneous segments and each entry of the indicator VSDHT$^2$ is associated with $d^{th}$ time interval is the $TD_d^{segmented}$ value for one segment.

Calculating the indicator VSDHT$^2$ requires historical data for the segmentation process. Therefore, we extracted the VSDHT$^2$ indicator from an initial interval $int_{start}$ corresponding to 5% of the average number of sample intervals of the training bearings lifetime. Before $int_{start}$, the conditions are considered as at the highest health status.

4.2.2 Cumulative Percent Variance (CPV)

In order to calculate the values of SDHT$^2$, we need to set the number $p$ of retained principal components.

$$CPV = \left( \frac{\sum_{g=1}^{p} \lambda_g}{\sum_{g=1}^{E} \lambda_g} \right) \times 100\% \tag{8}$$

where $\lambda_g$ denotes the variance of $g^{th}$ principal component.

When the principal components $\varphi$ and their variances $\lambda$ are sorted by variance $\lambda$ in descending order, CPV is the measurement of the percent variance extracted by the first $p$ principal components as defined in equation (8) (Valle et al., 1999).

The selected number $p$ of retained principal components is set such that the CPV value is equal at least to 95% for every bearings training sample.

4.2.3 Parameters of the Extraction Algorithm

The parameters of the extraction algorithm are $k$ and $m$.

Spearman rank correlation coefficient (SRCC) is a nonparametric technique for evaluating the degree of linear association or correlation between two independent variables (Best et al., 1975). We respectively calculate the SRCC between the time order series and the VSDHT$^2$ data series of each variable dimension from every training sample. We use the average standard deviation of SRCC (ASDS) defined in equation (9) to compare the VSDHT$^2$ performance for different values of parameters $k$ and $m$.

$$ASDS = 1/H \sum_{s=1}^{H} \sqrt{1/(m-1) \sum_{d=1}^{m} (SRCC_d^s - (1/m \sum_{d=1}^{m} SRCC_d^s))^2} \tag{9}$$

where $s$ is the label ($s=1, 2, ..., H$) of the training bearings sample, and $H$ is the number of training bearings samples.

The ASDS measures the divergence between different variables dimensions of VSDHT$^2$ data series associated to the amount of information carried by the indicator. The bigger value of ASDS, the more effective VSDHT$^2$. By comparing the ASDS values based on different $k$ and $m$ values, we choose the VSDHT$^2$ data series that carries the more information. Thus, we choose the values of $k$ and $m$ that correspond to a largest value of ASDS.

5 The NVSDHT$^2$ indicator

To generalize the indicator VSDHT$^2$, we need to normalize the indicator values. The indicator VSDHT$^2$ value corresponding to each observation is a vector. Each element is the segmented discarded Hotelling T square. Substituting (2) (5) into the formula (6), we can get (10).

$$TD_d^{segmented} = \frac{1}{B} \sum_{v=1}^{B} (\varphi_{d,v}^{1 \sim E} - \vartheta_d) \phi_d^{E^{-1}} (\varphi_{d,v}^{1 \sim E} - \vartheta_d)^T - \frac{1}{B} \sum_{v=1}^{B} (\varphi_{d,v}^{1 \sim p} - \vartheta_d) \phi_d^{p^{-1}} (\varphi_{d,v}^{1 \sim p} - \vartheta_d)^T \tag{10}$$

Where $\phi^E_d$ is covariance matrix of all the principal components. Since the principal components are linearly uncorrelated, all the off-diagonal elements of $\phi^E_d$ are small enough that one can consider them as 0. Hence $\phi^E_d$ can be approximated as diagonal matrix $\psi^E_d$ which has the entries of the principal diagonal in $\phi^E_d$ as its main diagonal. In the same way, $\phi^p_d$ can be approximated as diagonal matrix $\psi^p_d$ which has the entries of the principal diagonal in $\phi^p_d$ as its main diagonal. (10) can be rewritten as (11).

$$TD_d^{segmented} = \frac{1}{B}\sum_{v=1}^{B}(\varphi_{d,v}^{1\sim E} - \vartheta_d)\psi^E_d{}^{-1}(\varphi_{d,v}^{1\sim E} - \vartheta_d)^T - \frac{1}{B}\sum_{v=1}^{B}(\varphi_{d,v}^{1\sim p} - \vartheta_d)\psi^P_d{}^{-1}(\varphi_{d,v}^{1\sim p} - \vartheta_d)^T \quad (11)$$

In addition, all the entries of the principal diagonal in $\phi^E_d$ are the variances of two variables among the principal components. The diagonal matrix $\psi^E_d$ its non-zero entries are the variances of two variables among the principal components. According to the principle of linear algebra (Strang, 1993), $\psi^E_d{}^{-1}$ is a diagonal matrix which has the entries of the principal diagonal are the reciprocal of the entries of the principal diagonal in $\phi^E_d$. So, the item before the minus sign of the right side in formula (11) can approximately equal to $E$. In the same way, the item after the minus sign of the right side in formula (11) can approximately equal to $p$. So, the $TD_d^{segmented}$ values is near the value of ($E$-$p$). In this way, we can defined the indicator NVSDHT$^2$ as in (12).

$$NVSDHT^2 = \left[TD_1^{segmented}/(E-p), TD_2^{segmented}/(E-p), \cdots, TD_m^{segmented}/(E-p)\right]^T \quad (12)$$

Table I VSDHT$^2$ ASDS Value

| k~m | ASDS |
|---|---|
| 200~20 | 0.0330 |
| 200~10 | 0.0264 |
| 200~5 | 0.0152 |
| 100~20 | 0.0280 |
| 100~10 | 0.0411 |
| 100~5 | 0.0738 |
| 50~20 | 0.0186 |
| 50~10 | 0.0217 |
| 50~5 | 0.0221 |

**6 Numerical results**

We use the benchmark data introduced in Section 2 to test the effectiveness of three kinds of indicators we proposed for monitoring the bearing degradation.

The Figure 5 shows the indicator SDHT$^2$ for bearings 1-1 to 1-7.

After processing the selected common features $fc$ of the whole lifecycle of the training bearings under condition 1 (bearing1-1 and 1-2) by PCA, Figure 6 displays the cumulative percent variances associated with the number of principal components.

The results depicted on Figure 6 show that, for the training data, when the number $p$ of principal components is 2, the cumulative percent variance from bearing1-1 and 1-2 are both greater than 95%. Consequently, we choose $p$=2 for all the bearings under condition 1.

According to equation (9), the ASDS for VSDHT$^2$ with different $k$, $m$ values from the two training bearings 1-1 and 1-2 are respectively displayed in table I. So, in this work, we choose the $k$ as 100, $m$ as 5 for VSDHT$^2$.

The indicator VSDHT$^2$ data series from $int_{start}$ to the final time point for bearings 1-1 to 1-7 are displayed in Figure 7. The VSDHT$^2$ data series show monotonic trend during the bearing degradation and is sensitive to the whole bearing lifetime.

The indicator NVSDHT$^2$ for bearings 1-1 to 1-7 is showed in Figure 8.

# 7 Conclusion

In this paper, we proposed three new bearing degradation monitoring indicators SDHT$^2$, VSDHT$^2$ and NVSDHT$^2$ which all incorporate discarded space information. These indicators describe the whole degradation process history, and carry the real-time information of bearings degradation. Moreover, the NVSDHT$^2$ is the generalized version of VSDHT$^2$ and is promising to monitor bearings degradation.

The effectiveness of the indicators proposed in the literature was tested by numerical experiments using the IEEE PHM 2012 benchmark data. In general, the obtained three kinds of indicators all show monotonic trend during the bearing degradation and are all sensitive to the degradation all along the bearings lifetime.

Current work focuses on investigating appropriate methods to exploit the information provided by the three indicators for estimating the bearing remaining useful life.

**Figure 1** Parameters of Sample (Nectoux et al., 2012)

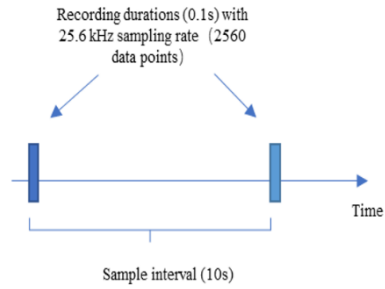

**Figure 2** Selected common feature RMS of bearing1-2

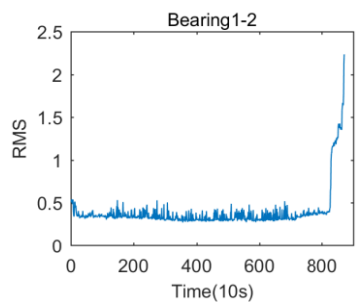

**Figure 3** The relationships among the original space and projected spaces through the PCA method

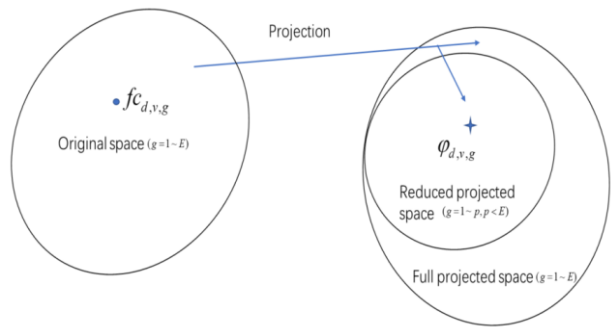

**Figure 4** Bottom-Up segmentation algorithm

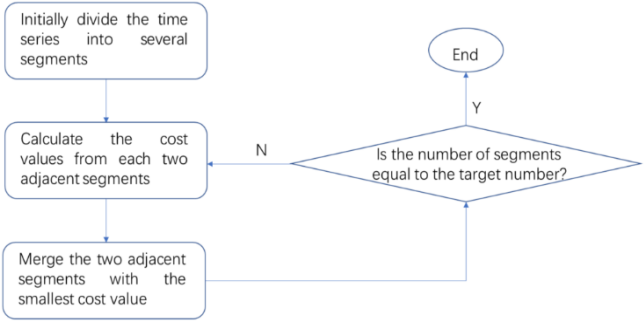

**Figure 5** SDHT$^2$ of 7 bearings under condition 1

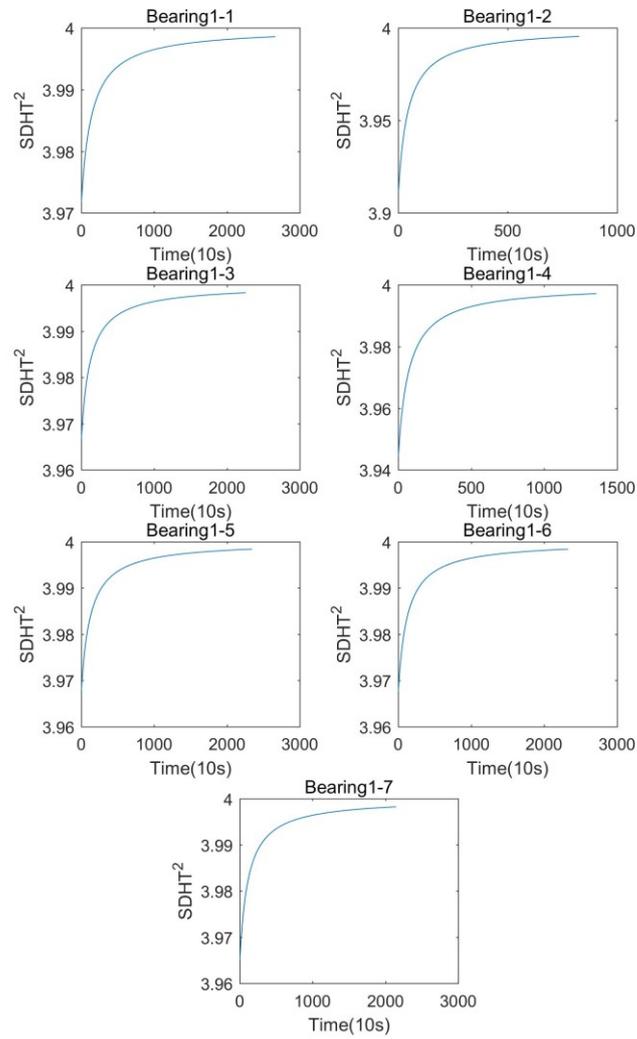

**Figure 6** Cumulative percent variance of bearing1-1 and 1-2

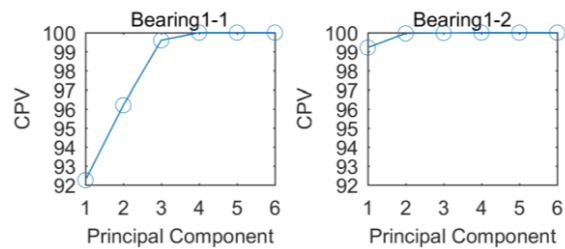

**Figure 7** VSDHT$^2$ of 7 bearings under condition 1

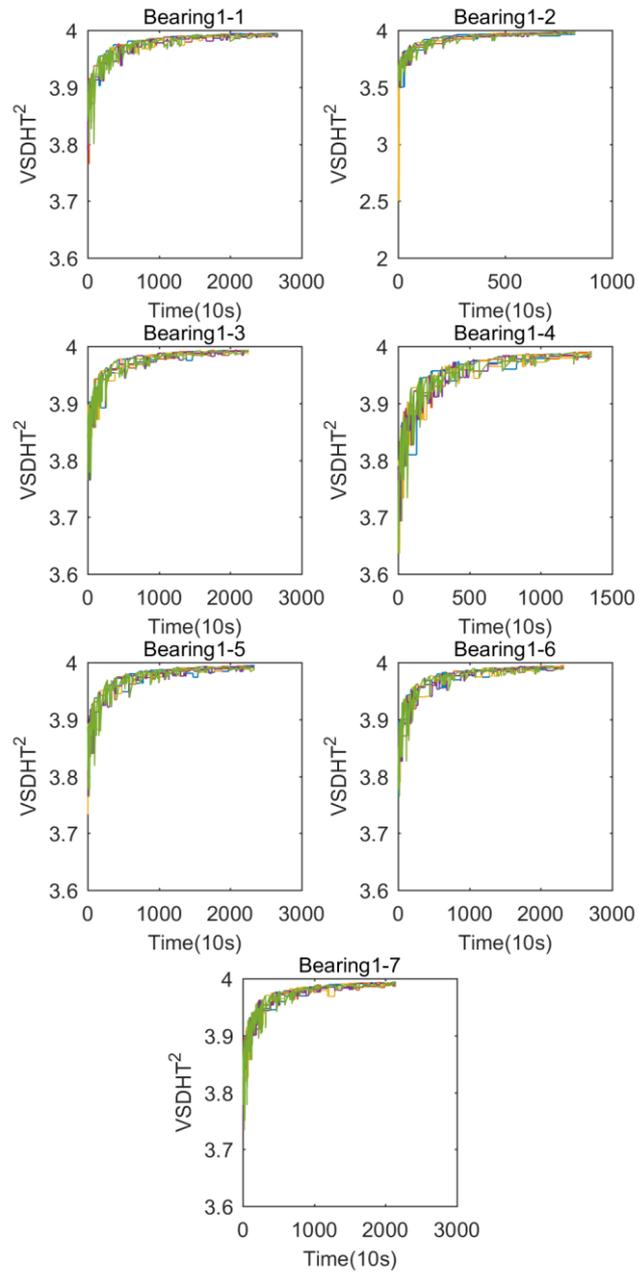

**Figure 8** NVSDHT$^2$ of 7 bearings under condition 1

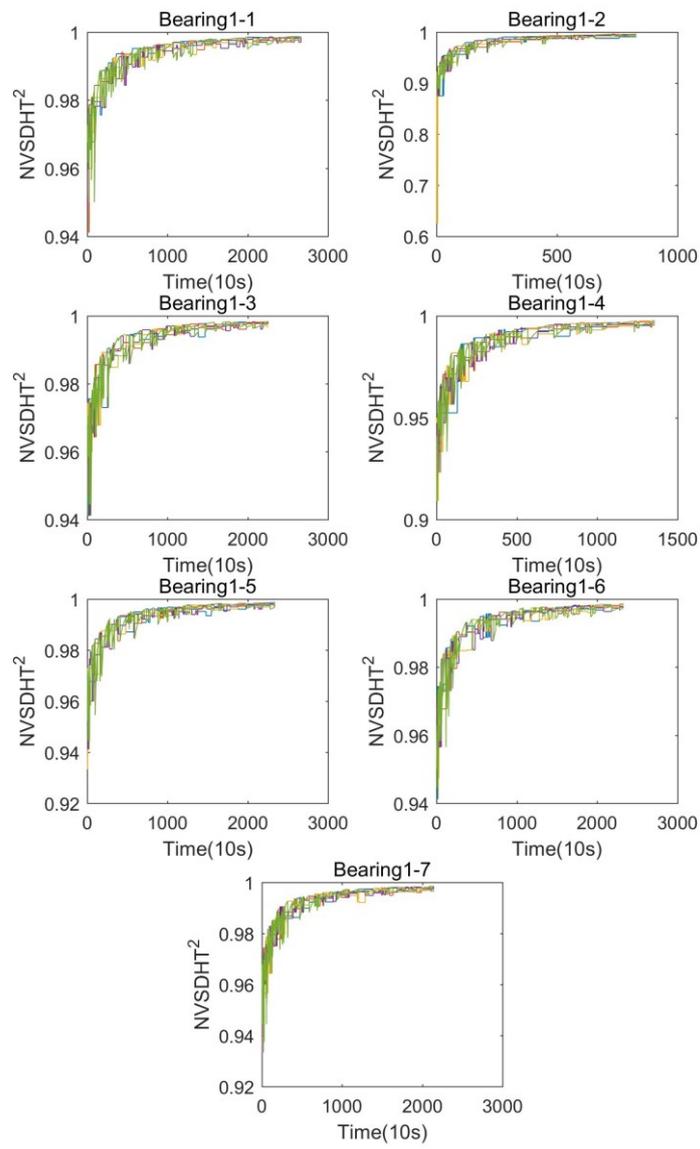